\documentclass[journal]{IEEEtran}
\usepackage{float}
\usepackage{graphicx}
\usepackage{subfigure}
\usepackage{epstopdf}
\usepackage{color}
\usepackage{amssymb}
\usepackage{amsmath}
\usepackage{algorithm}
\usepackage{algorithmic}
\usepackage{bm}
\usepackage[colorlinks,citecolor=blue,linkcolor=blue,urlcolor=blue]{hyperref}
\usepackage{stfloats}
\begin{document}

\title{\Large FAS-Driven Spectrum Sensing for Cognitive Radio Networks}
\author{Junteng Yao, 
            Ming Jin, 
            Tuo Wu, 
            Maged Elkashlan,
            Chau Yuen, \emph{Fellow, IEEE},\\
            Kai-Kit Wong, \emph{Fellow}, \emph{IEEE}, 
            George K. Karagiannidis, \emph{Fellow}, \emph{IEEE}, 
            and Hyundong Shin, \emph{Fellow}, \emph{IEEE}
\vspace{-9mm}

\thanks{J. Yao and M. Jin are with the faculty of Electrical Engineering and Computer Science, Ningbo University, Ningbo 315211, China (E-mail: $\rm\{yaojunteng, jinming\}@nbu.edu.cn$). T. Wu and C. Yuen are with the School of Electrical and Electronic Engineering, Nanyang Technological University, 639798, Singapore (E-mail: $\rm \{tuo.wu, chau.yuen\}@ntu.edu.sg$). M. Elkashlan is with the School of Electronic Engineering and Computer Science at Queen Mary University of London, London E1 4NS, U.K. (E-mail: $\rm maged.elkashlan@qmul.ac.uk$). K. K. Wong is with the Department of Electronic and Electrical Engineering, University College London, Torrington Place, WC1E 7JE, United Kingdom and also with the Department of Electronic Engineering, Kyung Hee University, Yongin-si, Gyeonggi-do 17104, Korea (E-mail: $\rm kai\text{-}kit.wong@ucl.ac.uk$). G. K. Karagiannidis is with the Department of Electrical and Computer Engineering, Aristotle University of Thessaloniki, 54124 Thessaloniki, Greece (E-mail: $\rm geokarag@auth.gr$). H. Shin is with the Department of Electronic Engineering, Kyung Hee University, Yongin-si, Gyeonggi-do 17104, Korea (E-mail: $\rm hshin@khu.ac.kr$).}
\thanks{\emph{Corresponding author: Tuo Wu.}}

}
\maketitle

\begin{abstract}
Cognitive radio (CR) networks face significant challenges in spectrum sensing, especially under spectrum scarcity. Fluid antenna systems (FAS) can offer an unorthodox solution due to their ability to dynamically adjust antenna positions for improved channel gain. In this letter, we study a FAS-driven CR setup where a secondary user (SU) adjusts the positions of fluid antennas to detect signals from the primary user (PU). We aim to maximize the detection probability under the constraints of the false alarm probability and the received beamforming of the SU. To address this problem, we first derive a closed-form expression for the optimal detection threshold and reformulate the problem to find its solution. Then an alternating optimization (AO) scheme is proposed to decompose the problem into several sub-problems, addressing both the received beamforming and the antenna positions at the SU. The beamforming subproblem is addressed using a closed-form solution, while the fluid antenna positions are solved by successive convex approximation (SCA). Simulation results reveal that the proposed algorithm provides significant improvements over traditional fixed-position antenna (FPA) schemes in terms of spectrum sensing performance.
\end{abstract}


\vspace{-2mm}
\section{Introduction}
\IEEEPARstart{I}{n Internet}-of-Things (IoT) networks, the sheer number of devices requires significant bandwidth, putting pressure on the available spectrum \cite{IYaqoob17}. To tackle this situation, cognitive radio (CR) has emerged as an attractive solution that encourages secondary users (SUs) to opportunistically access the spectrum resources assigned to licensed primary users (PUs) \cite{AGharib21}. To do this, SUs sense and identify spectrum holes. If a frequency band is found to be unused, SUs can use that band, alleviating the spectrum scarcity problem \cite{TJiang24,JGe24,HXie24}.

For reliable spectrum sensing, many algorithms have been proposed, such as energy-based detection \cite{JYao19}, matched filtering methods \cite{WMJang14}, eigenvalue analysis-based approaches \cite{YZeng09}, and so on. However, due to complex wireless scenarios, factors such as channel fading, multipath effects, and noise could significantly degrade the detection performance of these methods. To address these challenges, some effective approaches have been proposed, such as cooperative spectrum sensing, integration of multiple algorithms, and increasing the sensing time. However, these methods bring performance improvements at the cost of strict synchronization requirements, increasing complexity, and sacrificing communication time. Therefore, more effective spectrum sensing approaches must be sought.

Recently, advances in reconfigurable antenna technologies seem to offer a solution. Known as fluid antenna system (FAS), it enables flexibility in antenna shape and position, introducing a new degree of freedom (DoF) to the physical layer of wireless communications \cite{KKWong21,Wong-2022fcn,New-submit2024,Wang-aifas2024}. Since its emergence, it has been investigated in a wide range of applications, e.g. \cite{Wong-2022,Khammassi-2023,New-2023,Yao20242,New-twc2023,Xu-2024,10318134,Hao-2024,HXu24,JYao24,Ghadi-2024bc,Ghadi-2024cache,Ghadi-2024ris,Yao20241,CWang24,LZhou24}. Prototypes on FAS have also been reported to validate its promising performance \cite{Shen-tap_submit2024,Zhang-pFAS2024}. Presumably, CR can take advantage of the capability of FAS to improve the sensing performance, which motivates this work.

To this end, this letter adopts FAS to maximize the detection probability in CR networks, and introduces a FAS-enabled SU capable of optimizing the positions of fluid antennas to better detect PU signals. This approach focuses on maximizing the PU detection probability under the constraints of false alarm probability and receive beamforming. To solve this non-convex problem, we derive a novel closed-form solution for the optimal detection threshold, which allows for a more tractable formulation. We then propose an alternating optimization (AO) algorithm that decomposes the reformulated problem into two tractable sub-problems. Specifically, the receive beamforming sub-problem is solved analytically in closed form, while the successive convex approximation (SCA) method is applied to optimize the antenna positioning. Our simulation results show that the proposed algorithm significantly outperforms existing benchmarks, highlighting the novelty of our approach.

\vspace{-1mm}
\section{System Model}
\begin{figure}[t]
\centering
\includegraphics[width=3in]{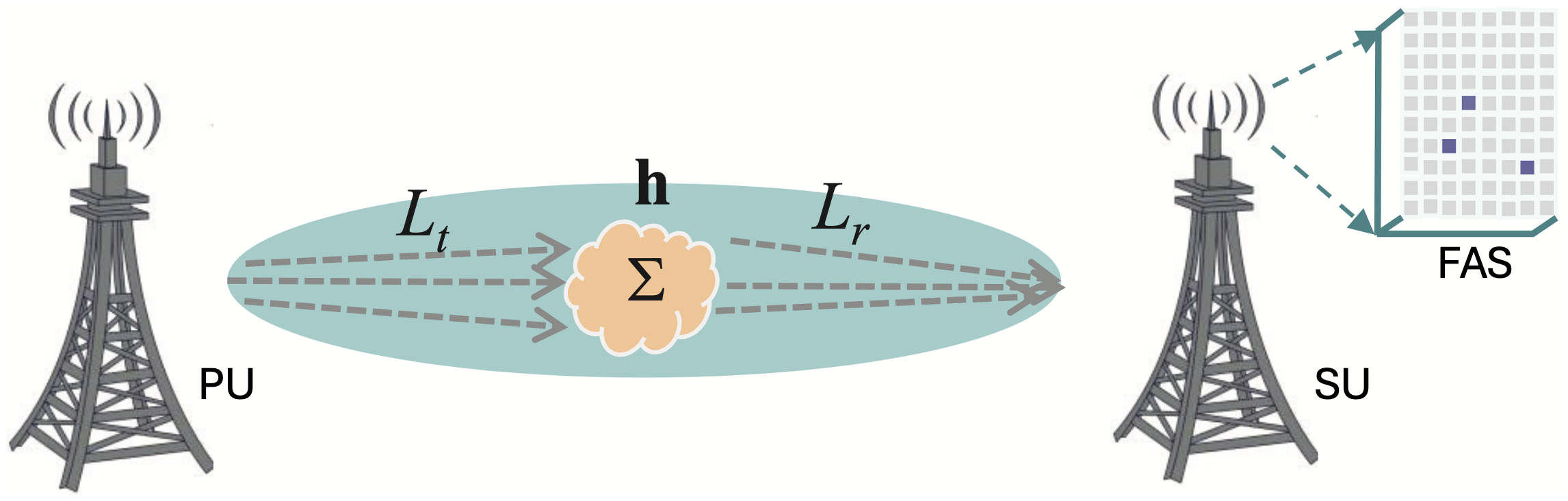}
\caption{FAS-driven spectrum sensing.}\label{system}
\vspace{-4mm}
\end{figure}

Consider the FAS-assisted CR system as depicted in Fig.~\ref{system}, where the PU is equipped with a conventional fixed-position antenna (FPA), while the SU employs $N (>1)$ fluid antennas. The SU determines if the PU signal is present through spectrum sensing and opportunistically accesses the PU's licensed spectrum if idle \cite{CWang24}. The SU's fluid antennas are connected to RF chains through integrated waveguides or flexible cables, and can be adjusted within a finite range $\mathcal{S}$ \cite{New-submit2024}.

\subsection{Channel Model}
We adopt the planar far-field model to express the channel. In this model, adjusting the fluid antenna positions changes the phase of path response, but does not affect the angles of arrival (AoAs), angles of departure (AoDs), or the amplitude of response coefficients for each path \cite{New-submit2024}. We assume that there are $L_t$ transmit paths and $L_r$receive paths in the PU-SU link, respectively. The position of the $n$-th fluid antenna is $\mathbf{t}_n=[x_n, y_n]^T$, and the set of SU antenna positions is represented as $\mathbf{\bar{t}} = [\mathbf{t}_1, \mathbf{t}_2,\dots, \mathbf{t}_N]$. The propagation distance difference between the $n$-th fluid antenna and the reference origin in the $l$-th receive path is given by $\rho_{l}(\mathbf{t}_n) = x_n\sin\theta_{l} \cos\varphi_{l} + y_n \cos\theta_{l}, l\in\{1,\dots,L_r\}$, where $\theta_{l} \in [0, \pi]$ and $\varphi_{l} \in [0, \pi]$ represent the elevation and azimuth angles of the $l$-th receive path, respectively. The far-field response vector of the fluid antenna for the PU-SU link can be written as
\begin{equation}\label{eq1}
\bm{\mathbf{f}}(\mathbf{t}_n)=\left [e^{j \frac{2\pi}{\lambda}\rho_{1}(\mathbf{t}_n)}, \dots, e^{j \frac{2\pi}{\lambda}\rho_{L_r}(\mathbf{t}_n)}\right]^T\in\mathbb{C}^{L_r\times 1},
\end{equation}
where $\lambda$ denotes the carrier wavelength. Therefore, the transmit field response matrix of the PU-SU link is given by
\begin{equation}\label{eq2}
\mathbf{F}\mathbf{(\bar{t})} =\left [\bm{\mathbf{f}}(\mathbf{t}_1),\bm{\mathbf{f}}(\mathbf{t}_2), \dots, \bm{\mathbf{f}}(\mathbf{t}_N)\right]\in\mathbb{C}^{L_r\times N}.
\end{equation}

On the other hand, since a single FPA is deployed at the PU, the transmit field response vector for the PU-SU link is simply $\mathbf{1}_{L_t}\in\mathbb{C}^{L_t\times 1}$. Also, we define the path response matrices $\bm{\Sigma}\in\mathbb{C}^{{L_r}\times {L_t}}$ as the path response of the PU-SU link. Consequently, the channel of the PU-SU link can be written as
\begin{equation}\label{eq3}
\mathbf{h}=\mathbf{F}\mathbf{(\bar{t})}^H\bm{\Sigma}\mathbf{1}_{L_t}\in\mathbb{C}^{N\times 1}.
\end{equation}

\vspace{-2mm}
\subsection{Signal Model}
To detect whether PU signals are present, we model spectrum sensing as a binary hypothesis problem. Thus, the $k$-th received signal sample at the SU is given by
\begin{equation}\label{eq4}
y(k)
=\left\{ \begin{array}{lc}
\mathbf{w}^H\mathbf{n}(k), &  \mathcal{H}_0,\\
\sqrt{P}\mathbf{w}^H\mathbf{h}x(k)+\mathbf{w}^H\mathbf{n}(k), & \mathcal{H}_1,
\end{array}\right.
\end{equation}
for $k \in\{1, \dots, K\}$, where $\mathcal{H}_0$ represents the null hypothesis that PU signals are not present; $\mathcal{H}_1$ is the alternative hypothesis to represent the absence of PU signals; and $\mathbf{w}\in\mathbb{C}^{N\times 1}$ with $\|\mathbf{w}\|^2=1$ is the receive beamforming of SU, which is used to equalize the received signals. Here, $\mathbf{n}(k) \sim \mathcal{CN}(\mathbf{0}, \sigma_n^2 \mathbf{I}_N)$ represents the additive circularly symmetric complex Gaussian (CSCG) noise, which is independent and identically distributed (i.i.d.) with mean $\mathbf{0}$ and covariance matrix $\sigma_n^2 \mathbf{I}_N$. The PU signal, denoted by $x(k)$, is also an i.i.d.~CSCG variable but with zero mean and unit variance, i.e., $x(k) \sim \mathcal{CN}(0, 1)$.

In the proposed FAS-driven spectrum sensing, we utilize energy detection to identify the PU signals. The received signal strength (RSS) at the SU during a sensing time slot serves as the test statistic, i.e.,
\begin{equation}\label{eq5}
T=\frac{1}{K}\sum_{k=1}^K|y(k)|^2.
\end{equation}

According to the central limit theorem, with a sufficiently large $K$, $T$ is approximately Gaussian distributed as \cite{HXie24}
\begin{equation}\label{eq6}
T\sim\left\{\begin{array}{lc}
\mathcal{N}\left(\sigma_n^2,\frac{\sigma_n^4}{K}\right), &  \mathcal{H}_0,\\
\mathcal{N}\left(\sigma_n^2(1+\gamma),\frac{\sigma_n^4(1+\gamma)^2}{K}\right), & \mathcal{H}_1,
\end{array}\right.
\end{equation}
where the signal-to-noise ratio (SNR) $\gamma$ is defined as
\begin{equation}\label{eq7}
\gamma = \frac{P|\mathbf{w}^H\mathbf{h}|^2}{\sigma_n^2}.
\end{equation}
Given a detection threshold $\lambda$, the SU determines whether the PU exists or not when $T>\tau$; otherwise, PU is deemed to be absent, which can be formulated by
\begin{align}\label{eq8}
T\overset{\mathcal{H}_1}{\underset{\mathcal{H}_1}{\gtreqless}}\tau.
\end{align}
Then, the false alarm probability and detection probability are, respectively, given by
\begin{align}
\label{eq9}\mathbb{P}_f&=Q\left(\frac{\tau-\sigma_n^2}{\sigma_n^2}\sqrt{K}\right),\\
\label{eq10}\mathbb{P}_d&=Q\left(\frac{\tau-\sigma_n^2(1+\gamma)}{\sigma_n^2(1+\gamma)}\sqrt{K}\right),
\end{align}
where $Q(z)$ is the Gaussian $Q$-function.

\vspace{-2mm}
\subsection{Problem Formulation}
Our objective is to maximize the detection probability of PU while satisfying the constraints of the false alarm probability and the fluid antennas' positions. Mathematically, we have
\begin{subequations}\label{eq12}
\begin{align}
\max\limits_{\mathbf{\bar{t}},\mathbf{w},\tau>0} \quad \ & \mathbb{P}_d \label{eq12a}\\
\mathrm{s.t.} \quad \ \  &\mathbb{P}_f \leq \delta, \label{eq12b}\\
&\mathbf{\overline{t}} \in \mathcal{S}, \label{eq12c}\\
&\|\mathbf{t}_n-\mathbf{t}_v\|_2\geq D,~n,v\in\mathcal{N},~n\neq v, \label{eq12d}\\
&\|\mathbf{w}\|_2^2=1, \label{eq12e}
\end{align}
\end{subequations}
where \eqref{eq12b} denotes the maximum false alarm probability constraint; \eqref{eq12d} is the minimum distance requirement between any two antennas with some prescribed minimum distance $D$ within the transmit region; and \eqref{eq12e} is the power constraint of the receive beamforming. Because of the non-convex objective function \eqref{eq12a} and constraints \eqref{eq12b} and \eqref{eq12d}, \eqref{eq12} is non-convex. We employ an AO algorithm to solve this problem.

\vspace{-2mm}
\section{The AO Algorithm}
In this section, we first reformulate Problem \eqref{eq12}, then use the AO algorithm to decompose Problem \eqref{eq12} into $N+1$ sub-problems, which are related to the receive beamforming and antennas' positions, and we obtain the locally optimal solution when  alternately optimizing these sub-problems.

Note that $Q(z)$ is a monotonically decreasing function with respect to $z$, and as a result, the objective function in Problem \eqref{eq12b} can be maximized when $\tau$ is minimized. Therefore, the constraint \eqref{eq12b} is active, i.e., $\mathbb{P}_f = \mathbb{P}_f^{\max}$. The optimal detection threshold is hence given by
\begin{align}\label{eq13}
\tau^o =\frac{\sigma_n^2Q^{-1}(\delta)}{\sqrt{K}}+\sigma_n^2.
\end{align}
Therefore, we can reformulate Problem \eqref{eq12}  as
\begin{align}\label{eq14}
\max\limits_{\mathbf{\bar{t}},\mathbf{w}} \  \mathbb{P}_d(\tau^o)~~{\rm s.t.}~~\eqref{eq12c}\text{--}\eqref{eq12e}.
\end{align}

\vspace{-2mm}
\subsection{Receive Beamforming Optimization}
Given $\mathbf{\bar{t}}$, Problem \eqref{eq12} can be rewritten as
\begin{align}\label{eq15}
\max\limits_{\mathbf{w}} \  \mathbb{P}_d~~{\rm s.t.}~~\eqref{eq12e}.
\end{align}
From \eqref{eq10}, $\mathbb{P}_d$ is a monotonically increasing function for $\gamma$. Thus, the optimal receive beamforming is given by
\begin{align}\label{eq16}
\mathbf{w}^o = \frac{\mathbf{h}}{\|\mathbf{h}\|}.
\end{align}

\vspace{-2mm}
\subsection{The $n$-th Fluid Antennas' Positions Optimization}
Given $\mathbf{w}$ and $\{\mathbf{t}_l\}_{l\neq n}$, by exploiting the monotonicity of the objective function, we can rewrite Problem \eqref{eq14} as
\begin{align}\label{eq18}
\max\limits_{\mathbf{t}_n} \  \gamma~~{\rm s.t.}~~\eqref{eq12c}, \eqref{eq12d}.
\end{align}
Though Problem \eqref{eq18} is non-convex, we employ SCA algorithm to solve it. First, we can rewrite $|\mathbf{w}^H\mathbf{h}|^2$ as
\begin{align}\label{eq21}
|\mathbf{w}^H\mathbf{h}|^2 &= \mathrm{Tr}\left(\mathbf{1}_{L_r}^H\bm{\Sigma}^H\mathbf{F}\mathbf{(\bar{t})}\mathbf{w} \mathbf{w}^H
\mathbf{F}^H\mathbf{(\bar{t})} \bm{\Sigma}_k \mathbf{1}_{L_r} \right)\nonumber\\
&=\alpha+\beta(\mathbf{t}_n)+2\mathbb{R}\{\mathbf{f}^H(\mathbf{t}_n) \bm{\Omega}\},
\end{align}
in which $\alpha=\mathrm{Tr}\left(\sum_{j\neq n}^{N} \mathbf{f}(\mathbf{t}_j)\mathbf{w}(j)\sum_{l\neq n}^{N} \mathbf{w}^H (l) \mathbf{f}^H(\mathbf{t}_l)\bm{\Phi} \right)$, $\beta(\mathbf{t}_n)=\mathrm{Tr}\left( \mathbf{w}(n)\mathbf{w}^H(n)\mathbf{f}_k(\mathbf{t}_n)\mathbf{f}^H(\mathbf{t}_n)\bm{\Phi}  \right)$, $\bm{\Phi}=\bm{\Sigma}\bm{\Sigma}^H$, and $\bm{\Omega}=\bm{\Phi}\left(\sum_{j\neq n}^{N} \mathbf{f}_k(\mathbf{t}_j)\right)\mathbf{w}^H(n)$.

Then we first obtain a lower bound for $\beta(\mathbf{t}_n)$ through the first-order Taylor expansion at point $\mathbf{f}(\mathbf{t}_n^{(m)})$, given by
\begin{align}\label{eq26}
2\mathbb{R}\{ \mathbf{f}^H(\mathbf{t}_n^{(m)})  \bm{\Psi} \mathbf{f}(\mathbf{t}_n) \}- \mathbf{f}^H(\mathbf{t}_n^{(m)}) \bm{\Psi} \mathbf{f}(\mathbf{t}_n^{(m)}),
\end{align}
where $\mathbf{t}_n^{(m)}$ is $\mathbf{t}_n$ at the $m$-th iteration, $\bm{\Psi}=\bm{\Phi}\mathbf{w}(n)\mathbf{w}^H(n)$. Now, we combine the third term in \eqref{eq21} with the first term in \eqref{eq26}, and thus have $\bar{\beta}(\mathbf{t}_n)=2\mathbb{R}\{\mathbf{f}^H(\mathbf{t}_n) \bm{\Upsilon}_n\}$,
where $\bm{\Upsilon}_n=\bm{\Psi}^H \mathbf{f}(\mathbf{t}_n^{(m)})+\bm{\Omega}$. We further use the second-order Taylor expansion to construct a global lower bound for $\bar{\beta}(\mathbf{t}_n)$ as
\begin{multline}\label{eq28}
g^l(\mathbf{t}_n) ={\bar{\beta}}(\mathbf{t}^{(m)}_n)+\nabla {\bar{\beta}}(\mathbf{t}_n^{(m)})^T\left(\mathbf{t}_n-\mathbf{t}_n^{(m)}\right)\\
-\frac{\kappa_n}{2}\left(\mathbf{t}_n-\mathbf{t}_n^{(m)}\right)^T\left(\mathbf{t}_n-\mathbf{t}_n^{(m)}\right),
\end{multline}
where
\begin{align}
\kappa_n=\frac{16\pi^2}{\lambda^2} \sum_{l=1}^{L_r} \lvert \bm{\Upsilon}_{l} \rvert
\end{align}
and $\nabla {\bar{\beta}}(\mathbf{t}_n)$ is the gradient vector of $\bar{\beta}(\mathbf{t}_n)$, with the detailed derivations given in Appendix \ref{appendixA}. Therefore, the concave lower bound of $|\mathbf{w}^H\mathbf{h}|^2$ can be found as
\begin{align}\label{eq29}
f^l\left(\mathbf{t}_n\right)=g^l(\mathbf{t}_n) + \alpha -\mathbf{f}^H(\mathbf{t}_n^{(m)}) \bm{\Psi}\mathbf{f}(\mathbf{t}_n^{(m)}).
\end{align}

For the constraint \eqref{eq12d}, we can relax $\|\mathbf{t}_n-\mathbf{t}_v\|_2$ to be a concave lower bound of $\mathbf{t}_n$ by using the first-order Taylor expansion at point $\mathbf{t}_n^{(m)}$, which is given by
\begin{align}\label{eq50}
\|\mathbf{t}_n-\mathbf{t}_v\|_2 \geq \frac{1}{\|\mathbf{t}_n^{(m)}-\mathbf{t}_v\|_2}(\mathbf{t}_n^{(m)}-\mathbf{t}_v)^T(\mathbf{t}_n-\mathbf{t}_v).
\end{align}
Then the constraint \eqref{eq12d} can be written as
\begin{align}\label{eq51}
\frac{1}{\|\mathbf{t}_n^{(m)}-\mathbf{t}_v\|_2}(\mathbf{t}_n^{(m)}-\mathbf{t}_v)^T(\mathbf{t}_n-\mathbf{t}_v) \geq D.
\end{align}

Accordingly, Problem \eqref{eq18} can be reformulated as
\begin{align}\label{eq52}
\max\limits_{\mathbf{t}_n} \ f^l\left(\mathbf{t}_n\right)~~{\rm s.t.}~~\eqref{eq12c}, \eqref{eq51},
\end{align}
which is convex, and can be solved using CVX \cite{MGrant}.

\vspace{-2mm}
\section{Simulation Results}
In this section, a two-dimensional (2D) coordinate system is considered in our simulations, where the PU and SU are located at $(0,0)$ m and $(250,0)$ m, respectively. We assume that the carrier frequency is $2.4$ GHz, which results in a wavelength $\lambda= 0.125$ m, and a minimum inter-antenna distance is $D=\frac{\lambda}{2}$. Also, we assume that the spatial region of fluid antennas at the SU is $\mathcal{S}=\left[-{\frac{A}{2}},{\frac{A}{2}}\right] \times \left[-{\frac{A}{2}},{\frac{A}{2}} \right]$, where $A=4\lambda$. We also assume that the number of transmit paths equals to the number of receive paths in the PU-SU link, i.e., $L_t=L_r=L=4$. Moreover, the path response matrices of all the links are modeled as $\bm{\Sigma}[l,l] \sim \mathcal{CN}(0,g_0 \left( d_k/d_0\right)^{-\alpha}/L), l=1,2,\dots, L$, where $d_k$ is the distance, $g_0=-40$ dB represents the average channel gain at the reference distance $d_0=1$ m and $\alpha=2.8$ is the path loss coefficient. The number of fluid antennas is $N=4$, and the transmit power of the PU is $P=10$ dBm. Besides, the noise power is $\sigma_n^2=-80$ dBm, the number of received signal samples is $K=1000$, and the maximum false alarm probability is set to $\delta=0.1$. The Monte Carlo simulations are conducted with $100$ time average, and the convergence accuracy of the proposed algorithm is $10^{-4}$.

\begin{figure*}[]
\centering
\begin{minipage}[b]{0.32\linewidth}
\centering
\includegraphics[width=2.5in]{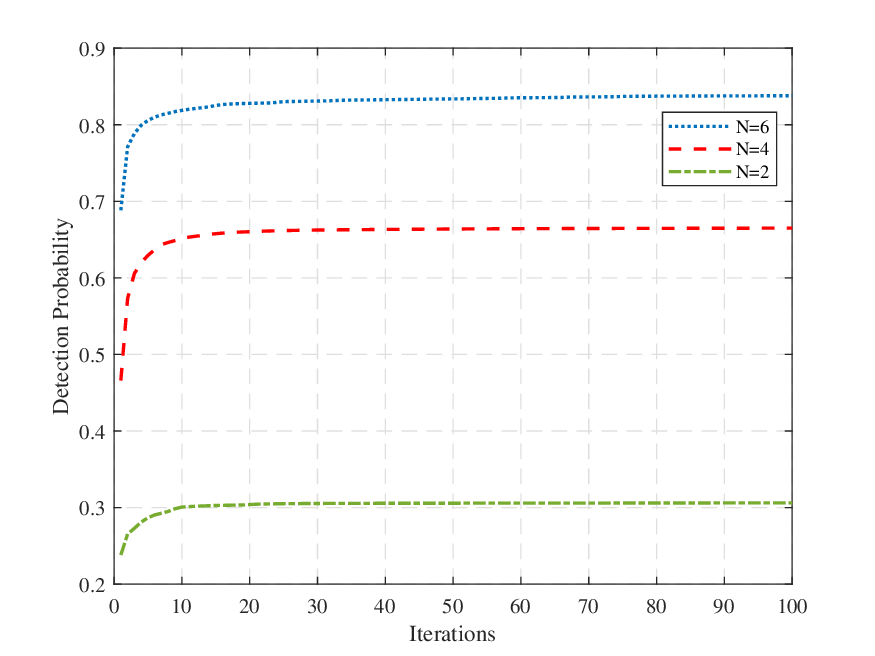}
\caption{Convergence of the proposed algorithm.}\label{iteration}
\end{minipage}
\hfill
\begin{minipage}[b]{0.32\linewidth}
\centering
\includegraphics[width=2.5in]{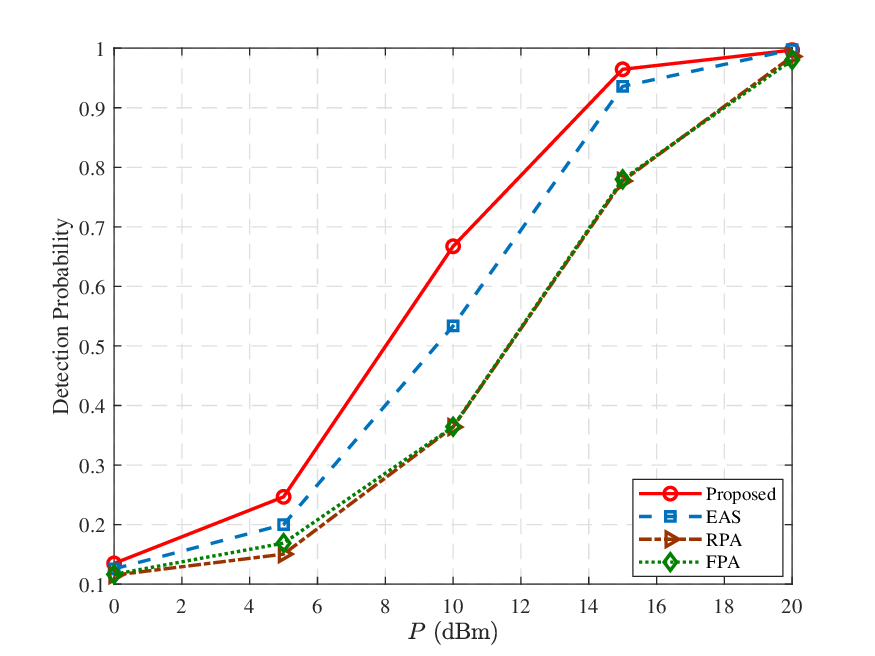}
\caption{The transmit power of the PU, $P$, versus the detection probability.}\label{DPvsP}
\end{minipage}
\hfill
\begin{minipage}[b]{0.32\linewidth}
\centering
\includegraphics[width=2.5in]{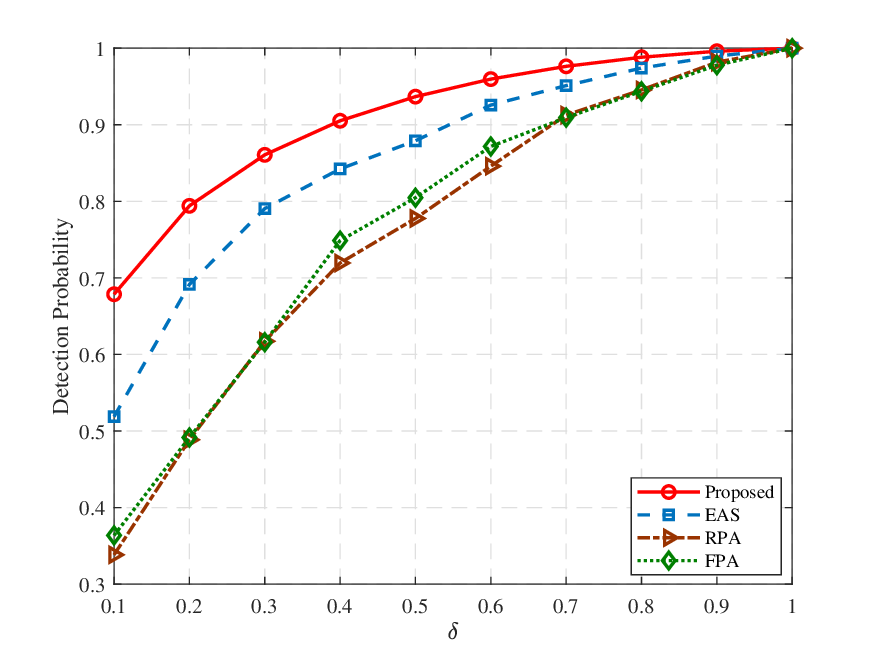}
\caption{The maximum false alarm probability of the SU, $\delta$, versus the detection probability.}\label{DPvsdelta}
\end{minipage} 
\vspace{-4mm}
\end{figure*}

In Fig.~\ref{iteration}, the results are provided to study the convergence performance of the proposed algorithm. As we can see, the AO algorithm converges with less than $30$ iterations for all three settings. Moreover, we can also observe that the detection probability increases with the increase in the number of fluid antennas at the SU. Also, the number of fluid antennas does not appear to affect the convergence speed much.

In Figs.~\ref{DPvsP} and \ref{DPvsdelta}, we compare the proposed FAS-assisted approach with several benchmarking methods: 
\begin{itemize}
\item $\mathbf{FPA}$, where $N$ FPAs are spaced at an interval of $\lambda/2$; 
\item $\mathbf{Random \ position \ antenna  \ (RPA)}$, in which $N$ antennas are randomly distributed within the spatial region $\mathcal{S}$, satisfying the constraint \eqref{eq12d}; and 
\item $\mathbf{Exhaustive\ antenna \ selection\ (EAS)}$, where $N$ antennas are selected from $2N$ fixed positions using an exhaustive search approach.
\end{itemize}

In Fig.~\ref{DPvsP}, we examine how the detection probability performance changes with the PU's transmission power. The results show that the proposed scheme has superior performance compared to all the benchmark methods. In addition, as the PU's transmit power \( P \) increases, the detection probability improves for all methods, with the proposed FAS scheme consistently outperforming the others, especially in the lower power range. \textit{This demonstrates its improved ability to detect the PU signal more reliably under varying power levels.}

Finally, we have the results in Fig.~\ref{DPvsdelta} to show the relationship between the maximum false alarm probability of the SU, denoted as \( \delta \), and the detection probability. The proposed FAS-based scheme shows a consistently higher probability of detection compared to the benchmark methods. As \( \delta \) increases, so does the probability of detection for all methods. However, the proposed scheme clearly outperforms the other approaches over the entire range of \( \delta \), especially in scenarios with lower false alarm probabilities. \textit{This indicates that the proposed method is robust in maintaining a high probability of detection while minimizing false alarms, thereby improving the overall performance of the CR system.}

\textbf{Remark 1}: \textit{The FAS scheme demonstrates superior performance in both PU and SU detection. It consistently outperforms benchmark methods,  particularly in challenging communication environments. This highlights the scheme's robustness in achieving high detection reliability for both PU and SU, thereby enhancing overall system efficiency.}

\vspace{-2mm}
\section{Conclusion}
In this letter, we studied a FAS-aided spectrum sensing in CR networks, in which an SU has the ability to optimize the positions of multiple fluid antennas to detect the signals from the PU. We maximized the detection probability at the SU by designing the receive beamforming and antennas' positions of the SU. Simulation results verified the significant impact of FAS on improving spectrum sensing performance.

\appendices
\section{Derivations of $\nabla {\bar{\beta}}(\mathbf{t}_n)$}\label{appendixA}
Given
\begin{align}\label{a1}
{\bar{\beta}_k}(\mathbf{t}_n)=2\left( \sum_{l=1 }^{L_r}   |\bm{\Upsilon}_l| \cos\left(\Pi_l(\mathbf{t}_n)\right)\right),
\end{align}
where $\Pi^{l}(\mathbf{t}_n)=\frac{2\pi}{\lambda}\rho_{l}(\mathbf{t}_n)-\angle\bm{\Upsilon}_l$, the gradient vector of ${\bar{\beta}}(\mathbf{t}_n)$ can be respresented as $ \nabla {\bar{\beta}}(\mathbf{t}_n)=\left[\frac{\partial {\bar{\beta}}(\mathbf{t}_n)}{\partial x_n},\frac{\partial {\bar{\beta}}(\mathbf{t}_n)}{\partial y_n}  \right]$, which can be, respectively, expressed as
\begin{align}
\frac{\partial{\bar{\beta}}(\mathbf{t}_n)}{\partial x_n} &=  -\frac{4\pi}{\lambda}\sum_{l=1}^{L_r} \lvert \bm{\Upsilon}_l \rvert\sin\theta_l\cos\varphi_l\sin(\Pi_l(\mathbf{t}_n)), \label{a3} \\
\frac{\partial{\bar{\beta}}(\mathbf{t}_n)}{\partial y_m} &=  -\frac{4\pi}{\lambda}\sum_{l=1}^{L_r} \lvert \bm{\Upsilon}_l \rvert \cos\phi_l\sin(\Pi_l(\mathbf{t}_n)).  \label{a4}
\end{align}

%\vspace{-2mm}

\end{document}